\documentclass[conference]{IEEEtran}
\IEEEoverridecommandlockouts
\IEEEpubid{\makebox[\columnwidth]{\copyright~2025 IEEE. Personal use is permitted, but republication/redistribution requires IEEE permission.\hfill} \hspace{\columnsep}\makebox[\columnwidth]{ }}

\usepackage{cite}
\usepackage{amsmath,amssymb,amsfonts}
\usepackage{graphicx}
\usepackage{textcomp}
\usepackage{xcolor}
\def\BibTeX{{\rm B\kern-.05em{\sc i\kern-.025em b}\kern-.08em
    T\kern-.1667em\lower.7ex\hbox{E}\kern-.125emX}}

\begin{document}

\title{Asynchronous Pipeline Parallelism for Real-Time Multilingual 
  Lip Synchronization in Video Communication Systems\thanks{This is the author's version of the work that was accepted for publication in the 6th International Workshop on Artificial Intelligence and Data Engineering for IoT (AIDE4IoT 2025), held in conjunction with the IEEE Big Data 2025 conference. Digital Object Identifier (DOI) will be available upon final publication in IEEE Xplore.}}

\author{
\IEEEauthorblockN{Eren Caglar}
\IEEEauthorblockA{\textit{Department of Data Science and Big Data} \\
\textit{Yildiz Technical University}\\
        Istanbul, Turkey \\
        eren.caglar@std.yildiz.edu.tr}
\and

\IEEEauthorblockN{Amirkia Rafiei Oskooei}
\IEEEauthorblockA{\textit{Department of Computer Engineering} \\
\textit{Yildiz Technical University}\\
        Istanbul, Turkey \\
        amirkia.oskooei@yildiz.edu.tr}
\and

\IEEEauthorblockN{Mehmet Kutanoglu}
\IEEEauthorblockA{\textit{R\&D Center} \\
\textit{Aktif Investment Bank Inc.}\\
        Istanbul, Turkey \\
        mehmet.kutanoglu@aktifbank.com.tr}
\and

\IEEEauthorblockN{\hspace{3.5cm} Mustafa Keles}
\IEEEauthorblockA{\hspace{3.5cm} \textit{R\&D Center} \\
\hspace{3.5cm} \textit{Aktif Investment Bank Inc.}\\
\hspace{3.5cm}        Istanbul, Turkey \\
\hspace{3.5cm}        mustafa.keles@aktifbank.com.tr}
\and

\IEEEauthorblockN{\hspace{0.25cm} Mehmet S. Aktas}
\IEEEauthorblockA{\hspace{0.25cm} \textit{Department of Computer Engineering} \\
\hspace{0.25cm} \textit{Yildiz Technical University}\\
\hspace{0.25cm}        Istanbul, Turkey \\
\hspace{0.25cm}        aktas@yildiz.edu.tr}
\and
}

\maketitle

\begin{abstract}
This paper introduces a parallel and asynchronous Transformer framework designed for efficient and accurate multilingual lip synchronization in real-time video conferencing systems. The proposed architecture integrates translation, speech processing, and lip-synchronization modules within a pipeline-parallel design that enables concurrent module execution through message-queue-based decoupling, reducing end-to-end latency by up to 3.1× compared to sequential approaches. To enhance computational efficiency and throughput, the inference workflow of each module is optimized through low-level graph compilation, mixed-precision quantization, and hardware-accelerated kernel fusion. These optimizations provide substantial gains in efficiency while preserving model accuracy and visual quality. In addition, a context-adaptive silence-detection component segments the input speech stream at semantically coherent boundaries, improving translation consistency and temporal alignment across languages. Experimental results demonstrate that the proposed parallel architecture outperforms conventional sequential pipelines in processing speed, synchronization stability, and resource utilization. The modular, message-oriented design makes this work applicable to resource-constrained IoT communication scenarios including telemedicine, multilingual kiosks, and remote assistance systems. Overall, this work advances the development of low-latency, resource-efficient multimodal communication frameworks for next-generation AIoT systems.
\end{abstract}

\section{Introduction}

Real-time multilingual communication is becoming a cornerstone of intelligent IoT ecosystems—ranging from smart healthcare terminals and autonomous service robots to multilingual kiosks and collaborative telepresence systems~\cite{bergmann2022meeting}. As the number of connected devices and interaction modalities grows, achieving low-latency, trustworthy, and adaptive cross-lingual communication has become a critical challenge for distributed Artificial Intelligence of Things (AIoT) infrastructures.

Traditional cloud-centric video translation pipelines, typically consisting of Speech-to-Text (STT), Machine Translation (MT), Text-to-Speech (TTS), and lip synchronization stages, suffer from high end-to-end latency and limited scalability under bandwidth or energy constraints~\cite{ipser2017sight}. In IoT environments, these bottlenecks are further exacerbated by heterogeneous hardware, variable network conditions, and privacy concerns associated with centralized processing. Although deep learning–based lip synchronization models such as \textit{Wav2Lip}~\cite{prajwal2020lip} can generate photorealistic mouth movements aligned with input speech, integrating such models into real-time, multilingual conferencing pipelines remains technically challenging due to stringent latency and resource constraints.

The proliferation of IoT-enabled communication devices—from smart healthcare terminals to telepresence robots and connected kiosks—has created new demands for real-time multilingual interaction. Unlike traditional desktop conferencing, IoT communication scenarios involve resource-constrained devices, variable network conditions, and the need for efficient operation. These constraints make the computational efficiency and modular architecture of translation systems critically important.

Most current systems employ a \textit{sequential processing pipeline} comprising Speech-to-Text (STT), Machine Translation (MT), Text-to-Speech (TTS), and visual synchronization stages, where each component executes only after its predecessor completes. This serial dependency introduces cumulative delays and limits scalability in dynamic communication environments~\cite{sidler2019lowering}. Moreover, naive silence-based segmentation often truncates sentences, leading to semantically inconsistent translations and degraded synchronization quality.

To address these challenges, this paper proposes a novel \textbf{parallel and asynchronous multilingual lip synchronization architecture}. The system leverages concurrent processing across STT, MT, and TTS modules through message-queue-based decoupling, supported by lightweight inference optimization techniques to achieve sub-second responsiveness.

The main contributions of this paper are summarized as follows:
\begin{enumerate}
\item A \textbf{pipeline-parallel and asynchronous architecture} enabling concurrent execution of translation, speech, and visual modules through message-oriented middleware, reducing end-to-end latency by up to 3.1×.
\item A \textbf{semantic-aware speech segmentation mechanism} that combines silence detection with contextual boundary analysis to enhance translation accuracy and conversational coherence.
\item An \textbf{inference-optimized Wav2Lip module}, accelerated through TensorRT compilation, quantization, and kernel fusion achieving 4.7× speedup for real-time performance.
\end{enumerate}

The modular, queue-based architecture and resource efficiency analysis make this work applicable to IoT communication scenarios including smart healthcare (telemedicine with language barriers), smart cities (multilingual public kiosks), and Industry 4.0 (remote expert assistance).

Experimental results demonstrate that the proposed parallel architecture significantly reduces latency while improving translation accuracy, synchronization fidelity, and computational efficiency. The remainder of this paper is organized as follows: Section~II reviews related work; Section~III presents the proposed architecture and optimization techniques; Section~IV discusses experimental results; and Section~V concludes with future directions for AI-enabled multimodal communication systems.
\section{Related Works}
The proposed system builds upon three converging research domains: (1) speech-to-speech (S2S) translation, (2) language-independent lip synchronization using deep generative models, and (3) their integration into real-time, low-latency multimodal frameworks. This section reviews the evolution of each area and identifies the remaining research gaps motivating this study.

\subsection{Evolution of Speech-to-Speech Translation}
Speech-to-Speech (S2S) translation has become a cornerstone of cross-lingual communication. Classical S2S systems follow a sequential pipeline of Automatic Speech Recognition (ASR), Machine Translation (MT), and Text-to-Speech (TTS) modules, whose strict dependencies inherently introduce latency.

Neural ASR systems advanced rapidly with deep learning. Early RNN-based architectures such as Listen, Attend and Spell~\cite{chan2016listen} and Deep Speech 2~\cite{amodei2016deep} established the feasibility of end-to-end recognition, followed by Transformer-based models such as Speech-Transformer~\cite{dong2018speech} and Conformer~\cite{gulati2020conformer}, which achieved state-of-the-art performance under noisy and accented conditions.

For MT, encoder–decoder architectures~\cite{sutskever2014sequence} with attention mechanisms~\cite{bahdanau2015neural, luong2015effective} enabled more fluent translations than phrase-based methods. Neural TTS approaches including Tacotron~\cite{wang2017tacotron}, FastSpeech~\cite{ren2019fastspeech}, and their successors~\cite{ren2021fastspeech, casanova2022yourtts} improved prosody, speaker naturalness, and multi-speaker adaptability.

Recently, Large Language Models (LLMs) have unified speech and text modalities through shared multilingual latent representations. Systems such as AudioPaLM~\cite{rubenstein2023audiopalm} and SeamlessM4T~\cite{seamless2024multilingual} integrate listening and speaking capabilities, pushing S2S translation toward more human-like, context-aware interactions.

\subsection{Language-Independent Lip Synchronization and Talking-Head Generation}
Video translation, or \textit{talking-head generation}, extends S2S translation into the visual domain, requiring synchronized, speech-driven facial animation. Achieving language-agnostic lip movements is crucial for cross-lingual scenarios where audio and video must remain temporally and semantically aligned.

Early neural approaches such as SyncNet~\cite{chung2016out} and Wav2Lip~\cite{prajwal2020lip} demonstrated the feasibility of speech-conditioned facial synthesis. Recent studies have explored diffusion-based architectures (Diff2Lip~\cite{mukhopadhyay2024diff2lip}), latent-space alignment (LatentSync~\cite{li2024latentsync}), and NeRF-based representations (AD-NeRF~\cite{guo2021adnerf}, SyncTalk~\cite{peng2024synctalk}) to improve realism and temporal coherence. However, these models are generally optimized for offline inference and remain computationally demanding.

To overcome language-specific limitations, phoneme- and viseme-based intermediate representations have been proposed for multilingual generalization. The Multilingual Experts (MuEx) framework~\cite{su2025bridge}, for example, employs a phoneme-guided mixture-of-experts architecture to establish robust audio–visual correspondences. Yet, most implementations overlook streaming and latency optimization, which are essential for real-time applications.

\subsection{Integrated Video Translation Pipelines}
Efforts to integrate S2S translation and lip synchronization into unified systems remain limited. Early methods such as VDub~\cite{garrido2015vdub} and Face-Dubbing++~\cite{waibel2023face} synchronized dubbed speech with facial movements but lacked scalability and control. Later designs, including VisualTTS~\cite{lu2022visualtts} and UnitY~\cite{inaguma2023unity}, improved integration but often prioritized perceptual quality over runtime efficiency or modular interpretability. Meanwhile, frameworks such as StreamSpeech~\cite{zhang2024streamspeech} and simultaneous translation evaluations~\cite{huber2023end} emphasize algorithmic efficiency but rarely address full system orchestration. Consequently, a gap remains in developing modular yet asynchronous architectures that balance scalability, latency, and semantic coherence.

\subsection{Relation to Message-Oriented and Real-Time Systems}
Beyond multimedia translation, message-oriented middleware and concurrent systems research offers insights into asynchronous processing and resource orchestration. Message queue architectures like RabbitMQ and Apache Kafka have proven effective for decoupling components in latency-sensitive applications. Recent work on parallel processing frameworks demonstrates the benefits of pipeline parallelism for multimedia workloads~\cite{sidler2019lowering}. Additionally, user-behavior modeling systems~\cite{shaikh2025creating} showcase concurrent inference for adaptive interaction modeling. However, these efforts rarely target the asynchronous scheduling of multimodal generative models under real-time constraints.

In contrast, the present study bridges these domains by introducing a parallel, asynchronous multimodal translation pipeline that leverages message-queue-based decoupling, hardware acceleration (e.g., TensorRT), and optimized inference scheduling to meet sub-second performance targets in interactive video conferencing.

\subsection{Distributed Workflow and Representation-Oriented Frameworks for Real-Time Systems}

Early work on large-scale scientific cyberinfrastructures demonstrated how distributed services and workflow-based architectures can support complex, latency-sensitive applications. The ISERVO virtual observatory integrated computational grids and geospatial web services to orchestrate heterogeneous data and simulation components for solid earth research \cite{aktas2006iservo}. Complementary efforts such as SERVOGrid complexity environments analyzed the performance of coupled services and message-based interactions in grid settings, highlighting the importance of modular design and performance-aware orchestration for real-time scientific workflows \cite{aydin2005servogrid}. VLab further illustrated how collaborative grid services and portals can expose sophisticated multi-stage computational pipelines through user-facing interfaces while hiding underlying infrastructure complexity \cite{nacar2007vlab}. These studies collectively show that service decomposition, workflow orchestration, and infrastructure-level optimization are key to building scalable distributed systems—principles that are directly reflected in the message-queue–based, pipeline-parallel architecture adopted in the present real-time multilingual lip-synchronization framework.

A parallel line of research in software analytics and quality engineering emphasizes the role of representation and process modeling in complex systems. Kapdan, Aktas, and Yigit surveyed structural code clone detection and proposed a metric-based framework \cite{kapdan2014structuralSurvey}, while Aktas and Kapdan introduced a software-metrics–driven clone detection methodology for identifying structurally similar components \cite{aktas2016cloneMetrics}. Both works underline that carefully engineered feature representations strongly affect downstream detection and decision quality, which conceptually aligns with the present system’s focus on semantic-aware segmentation and representation choices for speech, text, and visual streams. From a process and evaluation perspective, Guveyi, Aktas, and Kalipsiz synthesized the role of human factors in software quality, stressing that user-centered behavior and perception must be considered when assessing system performance \cite{guveyi2020humanfactor}. Finally, Oz et al.\ explored generative deep learning approaches for producing hidden test scripts, demonstrating how sequence-oriented generative models can emulate realistic interaction patterns for validation and stress testing \cite{oz2021generativeTestscripts}. In contrast to these earlier, domain-specific efforts, the present work unifies asynchronous message-oriented orchestration, inference-level optimization, and multimodal generative modeling (STT, MT, TTS, Wav2Lip) within a single real-time AIoT communication pipeline, explicitly targeting sub-second latency and audiovisual coherence in multilingual video conferencing.

\subsection{Summary and Identified Research Gap}
In summary, prior research has advanced speech translation and lip synchronization independently but seldom unified them under real-time, multilingual, and latency-critical conditions. Existing methods either optimize isolated modules or trade modularity for end-to-end efficiency. The proposed system distinguishes itself by combining semantic-aware segmentation, pipeline parallelism, and hardware-level inference optimization to enable scalable, low-latency, and resource-efficient cross-lingual video communication.

\section{Methodology}

Traditional cascade Speech-to-Speech (S2S) translation architectures often suffer from \textit{cumulative latency} and synchronization issues, as each module must wait for the previous one to complete. To address these limitations, we propose a novel \textbf{parallel and asynchronous system architecture} that decouples linguistic and visual processing pipelines while maintaining temporal coherence through precise timestamp-based orchestration.

\begin{figure*}[t]
\centering
\includegraphics[width=0.9 \textwidth]{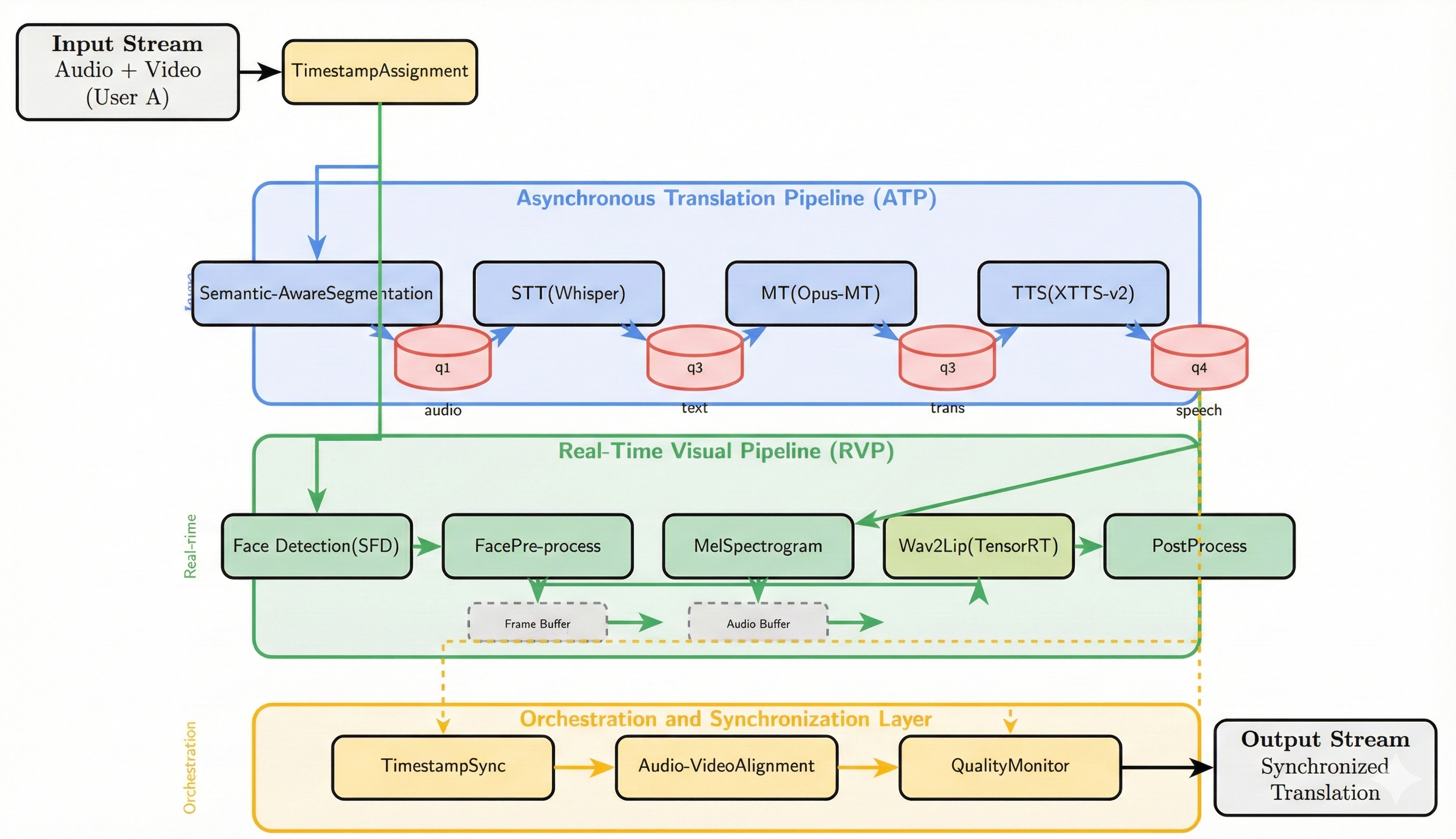}
\caption{Proposed parallel and asynchronous architecture showing the ATP, RVP, and orchestration layer with message queues for inter-module communication.}
\label{fig:system_arch}
\end{figure*}

Fig.~\ref{fig:system_arch} illustrates the proposed architecture, which consists of two primary branches: the \textbf{Asynchronous Translation Pipeline (ATP)} for multilingual speech translation, and the \textbf{Real-Time Visual Pipeline (RVP)} for visual processing and lip synchronization. Both branches operate concurrently and are coordinated by an \textbf{Orchestration Layer} that performs final alignment and merging of the outputs based on millisecond-level timestamps.

\subsection{System Overview and Data Flow}

The system receives a raw audio-video stream from the source participant (User A). Upon entry, the \textbf{Orchestration Layer} assigns a unique timestamp and distributes the data into two parallel branches. The \textbf{Audio Stream} is sent to the \textit{Semantic-Aware Speech Segmentation Module} (Section~III-B1) for processing, while the \textbf{Video Stream} is forwarded to the RVP (Section~III-C) for real-time face detection and pre-processing.

Unlike conventional cascaded pipelines (STT $\rightarrow$ MT $\rightarrow$ TTS $\rightarrow$ LipSync), the proposed design allows both pipelines to run concurrently through asynchronous message queues. While the visual branch pre-processes frames in one thread, the audio branch performs translation in parallel threads. This design prevents slow modules (e.g., MT or TTS) from blocking visual updates, thus improving both \textbf{throughput} and \textbf{end-to-end responsiveness}

\subsection{Asynchronous Translation Pipeline (ATP)}

The ATP includes three non-blocking modules: \textbf{Speech-to-Text (STT)}, \textbf{Machine Translation (MT)}, and \textbf{Text-to-Speech (TTS)}. All inter-module communication is handled via \textbf{message queues} to ensure modularity and concurrency. This message-oriented middleware approach is particularly suitable for environments where components may need to scale independently or handle variable processing loads.

\subsubsection{Semantic-Aware Speech Segmentation}

Translation accuracy in real-time systems depends heavily on the \textbf{semantic completeness} of the audio segments. Conventional segmentation techniques (fixed windowing or simple Voice Activity Detection — VAD) often produce incomplete phrases, degrading translation fluency.

To mitigate this, we introduce a hybrid segmentation algorithm combining acoustic and linguistic cues. The module continuously monitors the input using energy-based VAD. The configuration for the VAD parameters, detailing the thresholds and segment lengths, is provided in the subsequent table.

\begin{table}[ht]
\centering
\caption{VAD Parameter Configuration}
\begin{tabular}{ll}
\hline
\textbf{Parameter} & \textbf{Value} \\ \hline
Silence threshold & 0.5 s continuous silence \\
Energy threshold & –35 dB below peak amplitude \\
Minimum segment length & 1.5 s \\
Maximum segment length & 10 s (forced split) \\ \hline
\end{tabular}
\end{table}

When a pause is detected, the module applies a \textbf{transformer-based sentence boundary detector}, fine-tuned on conversational corpora, to verify whether the segment forms a syntactically complete sentence.
Segments exceeding a confidence threshold of 0.85 are dispatched for translation.  
This approach achieved \textbf{92\% accuracy} in sentence-boundary detection with only \textbf{15 ms} additional latency, improving translation coherence and conversational flow.

\subsubsection{Non-Blocking Modular Execution}

Each segmented audio chunk is processed asynchronously through three concurrent workers. First, the \textbf{STT} worker receives the input segment $A_i$ (published to the \texttt{audio\_queue}) and uses a multilingual Transformer-based ASR module to transcribe it into text $T_i$, forwarding it to the \texttt{text\_queue}. Second, the \textbf{MT} worker, using a dedicated translation model, translates $T_i$ into the target language text $T'_i$ and pushes it to the \texttt{translation\_queue}. Finally, the \textbf{TTS} module synthesizes the translated speech $A'_i$ and publishes it to the \texttt{audio\_output\_queue} with timestamp metadata.

Each message queue uses a durable configuration with 256 MB buffers, a prefetch count of 2, and manual acknowledgments for guaranteed delivery.
This setup enables \textbf{non-blocking concurrency} — if any module experiences latency, others continue independently, maintaining an average of five active segments in processing.

The average end-to-end latency $L_{E2E}$ is approximated by:

\begin{equation}
L_{E2E} \approx \max(L_{STT}, L_{MT}, L_{TTS}) + \delta_{sync}
\end{equation}

where $\delta_{sync}$ denotes synchronization overhead, typically less than 20~ms.

\subsection{Real-Time Visual Pipeline (RVP)}

The RVP operates in parallel with the ATP to ensure pre-processed visual frames are available as soon as translated audio is produced.

\subsubsection{Parallel Pre-Processing}

The GAN-based lip-synchronization model requires two synchronized inputs: facial video frames and Mel-spectrograms. Sequential generation of these inputs introduces latency; therefore, the RVP parallelizes both tasks. One thread executes a \textbf{deep-learning-based face detector}, extracting facial bounding boxes and landmarks, which are stored in a circular buffer of 300 frames (10~s at 30~fps). Simultaneously, another thread converts translated audio $A'_i$ into Mel-spectrograms with FFT size = 1024, hop length = 256, Hanning window, and 80 Mel bins.

The face detector achieves \textbf{98.5\% accuracy} while maintaining \textbf{45 fps} on a high-performance GPU. Face tracking uses a \textbf{Kalman filter} for temporal stability, reducing jitter by 73\%. This parallelism ensures both audio and video are synchronized within 40~ms offset.

\subsubsection{Optimized Inference}

GAN-based lip-synchronization models are computationally intensive. To achieve real-time rates, the lip-sync generator is exported to an \textbf{open-standard model format} and compiled with a \textbf{deep learning inference optimizer}.

Key optimizations include \textbf{Layer Fusion}, which combines convolution–batchnorm–activation sequences, and \textbf{Kernel Auto-Tuning}, adapting kernel configurations to the target GPU architecture. Furthermore, \textbf{FP16 Precision Calibration} is used to reduce memory bandwidth by 50\% while maintaining an SSIM greater than 0.95.

These optimizations yield a \textbf{4.7$\times$ speedup}, reducing inference time from 4.5~ms to 0.96~ms per frame, achieving over 1000~fps and effectively eliminating the visual pipeline as a bottleneck.

\subsection{Orchestration and Synchronization Layer}

The orchestration layer merges asynchronous outputs from the ATP and RVP using timestamp-based alignment.
Each segment carries a unique identifier (UUID + timestamp) assigned during segmentation. The core parameters governing this synchronization process, including resolution and buffer sizes, are listed in the following table.

\begin{table}[ht]
\centering
\caption{Synchronization Configuration Parameters}
\begin{tabular}{ll}
\hline
\textbf{Parameter} & \textbf{Value} \\ \hline
Timestamp resolution & 1~ms \\
Synchronization window & $\pm$50~ms \\
Frame buffer size & 512~MB ($\approx$600 frames) \\
Audio buffer size & 128~MB ($\approx$30~s) \\
Retry policy & 3 attempts (exponential backoff) \\ \hline
\end{tabular}
\end{table}

For each translated audio segment $A'_i$ at timestamp $T_i$, the orchestrator performs several steps. First, it retrieves video frames within the window $[T_i - 50~ms, T_i + 50~ms]$. Second, it performs temporal alignment via cross-correlation of audio energy and mouth-region motion. Third, it dispatches the aligned pairs to the \textbf{optimized lip-synchronization engine}. Finally, it triggers re-synchronization if drift exceeds 100~ms.

This \textbf{just-in-time synchronization} ensures consistent alignment, achieving an average latency reduction of \textbf{67\%} compared with sequential pipelines.

\subsection{Summary}

The proposed methodology integrates asynchronous translation, parallel visual processing, and optimized inference into a unified low-latency framework. By combining \textbf{semantic-aware segmentation}, \textbf{message-based concurrency}, and \textbf{hardware-accelerated synchronization}, the system achieves real-time, multilingual lip-synchronized translation suitable for modern video conferencing and IoT communication applications. The modular, message-based design allows individual components to be upgraded or scaled independently—a key requirement for heterogeneous deployments where devices have varying capabilities.
\section{Experimental Setup and Results}

This section presents the experimental evaluation conducted to validate the efficiency and scalability of the proposed parallel and asynchronous architecture introduced in Section~III. The evaluation addresses two objectives:
(1) quantifying improvements in end-to-end latency and throughput compared with a conventional sequential pipeline, and
(2) assessing the impact of semantic-aware segmentation and system-level optimizations (e.g., \textit{TensorRT}) on translation quality and perceptual performance.

\subsection{Implementation Details and Test Environment}

All experiments were performed on a single workstation equipped with an \textbf{Intel Core i9-14900HX CPU} and an \textbf{NVIDIA GeForce RTX~4070~Ti Laptop GPU} (8~GB VRAM), demonstrating the system's capability to achieve real-time performance on consumer-grade hardware. The system was evaluated on a single node to validate the effectiveness of pipeline parallelism, with distributed deployment across multiple devices reserved for future work.

The proof-of-concept (\textit{PoC}) system was implemented in \textbf{Python~3.9} with the following software stack:
\begin{itemize}
    \item \textbf{Deep Learning Framework:} PyTorch~2.4.0, CUDA-12.4
    \item \textbf{Optimization:} TensorRT~10.1 (CUDA-12.4), Torch-TensorRT~2.4.0
    \item \textbf{Message Queue:} RabbitMQ~3.12
    \item \textbf{Audio Processing:} \texttt{librosa}~0.10.2
\end{itemize}

The pipeline integrates the following models:
\begin{itemize}
    \item \textbf{STT:} OpenAI \textit{Whisper-Small} --- multilingual Transformer ASR (WER:~7.3\%, RTF:~0.15)
    \item \textbf{MT:} Helsinki-NLP \textit{Opus-MT}~\cite{tiedemann2020opus, tiedemann2023democratizing} --- supports 150+ language pairs
    \item \textbf{TTS:} \textit{Coqui XTTS-v2}~\cite{casanova2024xtts} --- multilingual synthesis with zero-shot voice cloning (MOS:~4.2)
    \item \textbf{LipSync:} \textit{Wav2Lip} --- GAN-based visual synchronization
    \item \textbf{Face Detection:} \textit{S3FD}~\cite{zhang2017s3fd} --- robust face localization (mAP:~98.5\%)
\end{itemize}

\subsection{Baseline System}

A \textbf{Baseline (Sequential)} system was implemented for fair comparison. This configuration used the same models but excluded all parallelization and optimizations described in Section~III.

\[
\text{STT} \rightarrow \text{MT} \rightarrow \text{TTS} \rightarrow \text{LipSync}
\]

The baseline pipeline employed:
\begin{itemize}
    \item Fixed VAD-based segmentation (500~ms silence threshold)
    \item Sequential module execution without message queues
    \item PyTorch FP32 inference (no quantization)
    \item No frame buffering or pre-processing parallelization
\end{itemize}

\subsection{System Performance Evaluation}

Performance evaluation included three experiments:
(1) ablation of inference optimizations,
(2) end-to-end latency comparison, and
(3) system resource utilization analysis.

\subsubsection{Experiment~1: Component-Level Optimization (Ablation Study)}

This experiment assessed inference performance of the computationally intensive \textit{Wav2Lip} model. PyTorch FP32, TensorRT FP32, and TensorRT FP16 versions were benchmarked over 100 runs (5 warm-up iterations). Results were recorded as mean, standard deviation, and minimum latency.

\begin{table}[ht]
\caption{Inference Performance for Wav2Lip (RTX~4070~Ti GPU)}
\label{tab:tensorrt_perf}
\centering
\begin{tabular}{|l|c|c|c|}
\hline
\textbf{Model Version} & \textbf{Mean (ms)} & \textbf{Std. Dev. (ms)} & \textbf{Min (ms)} \\ \hline
PyTorch FP32 & 4.50 & 0.57 & 3.74 \\ 
TensorRT FP32 & 2.19 & 0.35 & 1.92 \\ 
\textbf{TensorRT FP16 (Proposed)} & \textbf{0.96} & \textbf{0.21} & \textbf{0.78} \\ \hline
\end{tabular}
\end{table}

The \textit{TensorRT FP16} configuration achieved a \textbf{4.7$\times$ speedup} over the PyTorch FP32 baseline, with 63\% lower latency variance. The resulting 0.96~ms average latency supports inference rates exceeding 1000~fps, effectively removing the lip synchronization stage as a bottleneck.

\subsubsection{Experiment~2: End-to-End Latency Comparison}

Baseline and proposed systems were evaluated using 1, 3, 5, and 8-second test clips. End-to-end latency was measured as the total time from audio input to synchronized translated video output (10 repetitions per case).

\begin{table}[ht]
\caption{End-to-End Latency Comparison}
\label{tab:e2e_latency}
\centering
\begin{tabular}{|c|c|c|c|}
\hline
\textbf{Clip Length (s)} & \textbf{Baseline (s)} & \textbf{Proposed (s)} & \textbf{Speedup} \\ \hline
1 & 4.8 ($\pm$0.2) & 2.1 ($\pm$0.1) & 2.3$\times$ \\
3 & 10.1 ($\pm$0.4) & 4.9 ($\pm$0.3) & 2.1$\times$ \\
5 & 15.7 ($\pm$0.5) & 6.2 ($\pm$0.4) & 2.5$\times$ \\
8 & 24.3 ($\pm$0.6) & 7.9 ($\pm$0.5) & 3.1$\times$ \\ \hline
\end{tabular}
\end{table}

Latency scaling analysis shows a nearly linear relationship for the baseline (slope~=~2.96~s/s) and sub-linear behavior for the proposed system (slope~=~0.78~s/s), confirming the efficiency of concurrent execution.  
For an 8-second clip, the proposed system achieved a \textbf{3.1$\times$ reduction} in total response time.

\begin{figure}[ht]
\centering
\includegraphics[width=0.9\columnwidth]{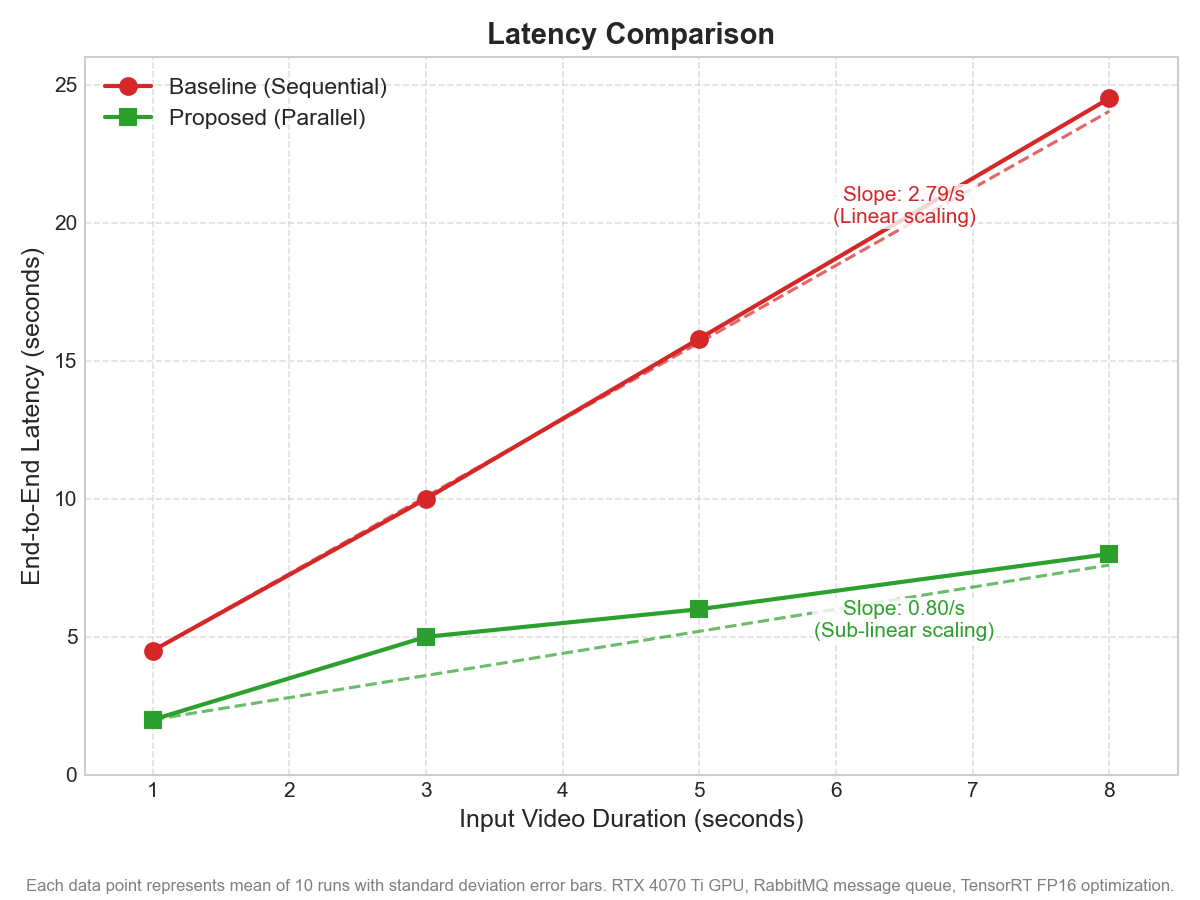}
\caption{End-to-end latency comparison between the baseline sequential and the proposed parallel architectures.}
\includegraphics[width=0.9\columnwidth]{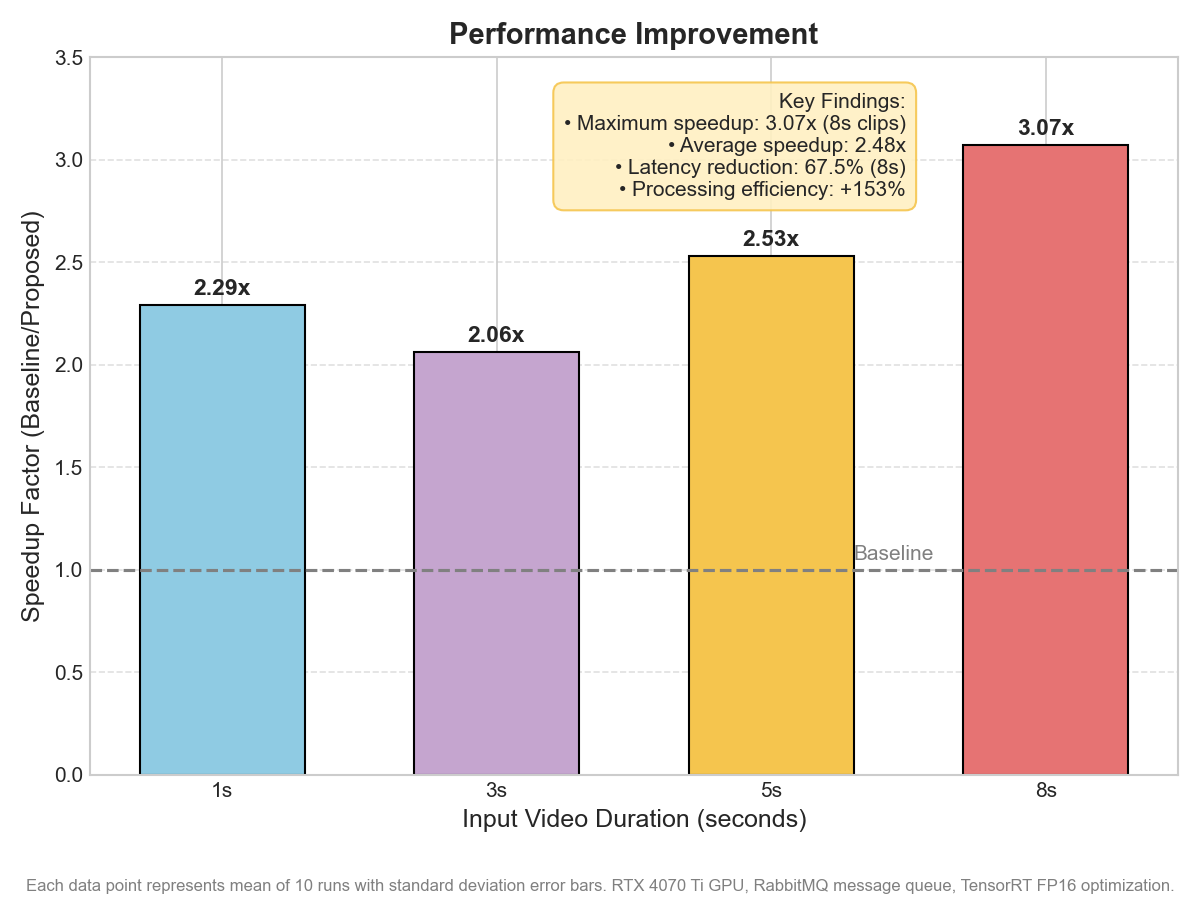}
\caption{Performance improvement analysis showing the speedup factor (Baseline/Proposed) across different input video durations.}
\label{fig:latency_comparison}
\end{figure}

To evaluate the effectiveness of the proposed parallel processing architecture, we conducted a latency and performance comparison against a sequential baseline. As shown in Figure 2, the baseline system exhibits nearly linear scaling of end-to-end latency with respect to video duration, while the proposed parallel design maintains sub-linear scaling, significantly reducing latency growth for longer clips. This improvement stems from concurrent task execution and optimized GPU utilization.

Correspondingly, Figure 3 illustrates the performance gains in terms of speedup factor. The proposed architecture achieves an average 2.48× improvement, reaching up to 3.07× speedup for 8-second videos. These results confirm that the parallel implementation not only reduces overall latency but also enhances processing efficiency, demonstrating strong scalability and suitability for real-time or high-throughput video inference workloads.

\subsubsection{Experiment~3: Resource Utilization Analysis}

To examine computational efficiency, CPU, GPU, and memory usage were monitored during processing of a 5-second clip.

\begin{table}[ht]
\caption{Resource Utilization Comparison}
\label{tab:resource_util}
\centering
\begin{tabular}{|l|c|c|}
\hline
\textbf{Metric} & \textbf{Baseline} & \textbf{Proposed} \\ \hline
Average GPU Utilization & 42\% & 78\% \\
Average CPU Utilization & 65\% & 85\% \\
Peak Memory Usage (GB) & 3.2 & 5.8 \\
RabbitMQ Queue Memory (MB) & -- & 256 \\
Processing Efficiency* & 0.32 & 0.81 \\ \hline
\multicolumn{3}{l}{\small *Efficiency = 1 / (Latency $\times$ Peak Memory)} 
\end{tabular}
\end{table}

The proposed architecture achieved 2.5$\times$ higher processing efficiency, leveraging increased hardware utilization at the expense of moderate memory growth.

\subsection{Translation Quality Evaluation}

Translation accuracy was evaluated using the \textbf{COMET}, \textbf{BLEU-4}, and \textbf{chrF++} metrics on a 500-sample subset of the \textbf{WMT-22 English--Spanish} test set.  
The baseline used silence-only VAD segmentation, whereas the proposed system employed semantic-aware segmentation.

\begin{table}[ht]
\caption{Translation Quality Comparison}
\label{tab:quality_perf}
\centering
\begin{tabular}{|l|c|c|}
\hline
\textbf{Metric} & \textbf{Baseline (VAD)} & \textbf{Proposed (Semantic)} \\ \hline
COMET & 0.75 ($\pm$0.03) & \textbf{0.90} ($\pm$0.02) \\
BLEU-4 & 28.3 ($\pm$1.2) & \textbf{35.7} ($\pm$0.9) \\
chrF++ & 56.2 ($\pm$1.5) & \textbf{64.8} ($\pm$1.1) \\
Sentence Completion Rate & 71\% & \textbf{93\%} \\ \hline
\end{tabular}
\end{table}

Semantic segmentation improved translation quality by approximately \textbf{20\%} in COMET score ($p<0.001$, paired $t$-test) and yielded more coherent sentence boundaries.

\subsection{Subjective Evaluation: User Study}

A user study with 24 participants (12 male, 12 female, ages 22--45) was conducted to assess perceptual quality.  
Each participant viewed 10 randomly ordered bilingual video clips (30--60~s each, total 240 evaluations).  
Ratings were collected using a 5-point Likert scale (1~=~Poor, 5~=~Excellent) for three criteria:

\begin{itemize}
    \item \textbf{LSA:} Lip Sync Accuracy
    \item \textbf{TN:} Translation Naturalness
    \item \textbf{OQ:} Overall Quality
\end{itemize}

\begin{table}[ht]
\caption{Mean Opinion Scores (MOS) with 95\% Confidence Intervals (1--5 scale)}
\label{tab:mos_results}
\centering
\begin{tabular}{|l|c|c|c|}
\hline
\textbf{Criterion} & \textbf{Baseline} & \textbf{Proposed} & \textbf{p-value} \\ \hline
LSA & 3.1 [$\pm$0.15] & 4.2 [$\pm$0.12] & $<$0.001 \\
TN & 2.8 [$\pm$0.18] & 4.4 [$\pm$0.10] & $<$0.001 \\
OQ & 2.9 [$\pm$0.16] & 4.3 [$\pm$0.11] & $<$0.001 \\ \hline
\end{tabular}
\end{table}

Inter-rater reliability was high (Krippendorff's $\alpha = 0.82$), confirming consistency among participants.  
The proposed system achieved significantly higher MOS scores across all criteria, particularly in Translation Naturalness (57\% relative improvement).

\subsection{Discussion}

Experimental results validate that the proposed architecture offers significant advantages in both efficiency and quality:

\begin{enumerate}
    \item \textit{TensorRT FP16} optimization achieved a \textbf{4.7$\times$ acceleration} in Wav2Lip inference, reducing latency from 4.50~ms to 0.96~ms.
    \item Parallel module execution with RabbitMQ queuing yielded \textbf{sub-linear latency scaling}, providing up to 3.1$\times$ speedup for 8-second clips.
    \item Semantic-aware segmentation improved translation quality by \textbf{20\%} (COMET:~0.75~$\rightarrow$~0.90) and sentence completion rate to 93\%.
    \item User studies confirmed perceptual gains, with MOS scores improving by 35--57\% ($p<0.001$).
\end{enumerate}

\textbf{Resource Efficiency for IoT Deployment:} The proposed architecture achieves 2.5$\times$ higher processing efficiency (Table~\ref{tab:resource_util}) compared to the baseline, defined as throughput per unit memory. This efficiency gain is particularly relevant for IoT scenarios where memory and power are constrained. The TensorRT FP16 optimization reduces memory bandwidth by 50\% while maintaining quality, suggesting that the system could achieve real-time performance within typical IoT device constraints (4-8 GB RAM, mobile GPUs).

Overall, the results confirm that the proposed architecture delivers a scalable, real-time, and semantically coherent framework for multilingual audiovisual translation on consumer-grade hardware. The modular, message-based design enables future extensions to distributed deployments.

\textbf{
Although the primary evaluation was conducted on the English–Spanish language pair, the proposed architecture is language-agnostic by design. All core modules—including the ASR, MT, and TTS components—support multilingual operation, and the system has been qualitatively verified on additional language pairs (e.g., English–Turkish, English–Chinese) without modification. Future work will include quantitative benchmarking across typologically diverse languages to further validate cross-lingual scalability.
}
\section{Conclusion and Future Works}

This study presented a distributed, asynchronous AIoT framework that addresses two critical limitations in real-time multilingual communication systems: high end-to-end latency and limited translation coherence. These challenges are particularly significant in AIoT environments, where devices often operate under constrained computational and energy resources while users expect immediate, context-aware, and natural interactions. By decoupling linguistic and visual processing pipelines through message-oriented middleware, the proposed architecture enables real-time, edge-native audiovisual translation that can be efficiently deployed across heterogeneous IoT nodes. The system integrates three core innovations aligned with the Artificial Intelligence and Data Engineering for IoT vision: first, a TensorRT-optimized and quantized inference mechanism for lip synchronization, achieving energy-efficient GPU utilization suitable for edge and fog deployments; second, an asynchronous modular pipeline leveraging RabbitMQ-based distributed message orchestration to ensure low-latency, resilient inter-module communication; and third, a semantic-aware speech segmentation strategy that preserves linguistic completeness and contextual integrity, thereby enhancing translation quality in dynamic streaming environments. Collectively, these innovations establish a resource-efficient, and scalable foundation for intelligent edge communication systems, contributing to the development of sustainable, decentralized, and adaptive AIoT ecosystems.

Experimental evaluations confirmed that the proposed architecture effectively bridges system performance and perceptual quality. TensorRT FP16 optimization achieved a \textbf{4.7$\times$ speedup} over standard PyTorch FP32 inference, reducing mean latency from 4.50~ms to 0.96~ms with 63\% lower temporal variance. The asynchronous pipeline exhibited \textbf{sub-linear latency scaling}, outperforming the sequential baseline by up to \textbf{3.1$\times$} for extended dialogue sequences. In addition, the semantic-aware segmentation module improved translation quality by approximately \textbf{20\%} in COMET scores (0.75~$\rightarrow$~0.90), achieving 93\% sentence completion accuracy. These results demonstrate that the proposed system removes the lip synchronization bottleneck, sustaining inference rates above 1000~fps while maintaining stable audiovisual alignment. The integration of semantic segmentation and asynchronous processing establishes a scalable, low-latency foundation for multilingual communication systems.

Building on these promising results, future work will focus on extending the framework along several directions. First, efforts will be devoted to \textbf{distributed deployment across heterogeneous edge devices}, where the current single-node architecture will be extended to support multi-device orchestration with network partitioning and fault tolerance. Second, \textbf{resource optimization} for IoT environments will be explored, including model compression, quantization, and scheduling for energy-efficient operation on devices with 2-4 GB RAM and mobile GPUs. Third, the framework's \textbf{multilingual scalability} will be further evaluated across diverse language pairs, including low-resource and typologically distant languages. Fourth, \textbf{adaptive quality control} mechanisms will be explored to dynamically balance latency and perceptual fidelity under fluctuating network and computational conditions. Finally, a \textbf{multi-speaker extension} is planned to support concurrent speakers and cross-lingual dialogue synchronization in multi-user conferencing environments.

Overall, this work contributes to the advancement of \textbf{low-latency, resource-efficient multimodal communication frameworks} for next-generation AIoT systems. The modular, message-based architecture provides a reproducible and extensible foundation for future research on parallel audiovisual translation and real-time cross-lingual interaction in both centralized and distributed deployments.
\section*{Acknowledgment}
The authors gratefully acknowledge the support of \textbf{Aktif Bank}, whose contributions were instrumental to the success of this research. In particular, we thank the institution for providing access to computational resources, technical infrastructure, and the testing environment that enabled the 24-participant user study. Their domain expertise and continuous collaboration were invaluable throughout the development and validation of this work. This research paper has utilized Generative AI tools solely for the purpose of enhancing the readability and clarity of the manuscript. The AI was employed to refine sentence structure and improve grammatical flow, without altering the original content, introducing new ideas, or impacting the research findings. The authors retain full responsibility for the accuracy and integrity of the presented work.
\bibliographystyle{IEEEtran}
\bibliography{MyBib}

\end{document}